\documentclass[final,1p,times,twocolumn]{elsarticle}
\usepackage{epsfig}
\usepackage{amssymb}
\usepackage{amsmath}
\journal{Nuclear Physics B}
\usepackage{xcolor}

\DeclareMathAlphabet{\oldcal}{OMS}{cmsy}{m}{n}
\newcommand{\bigo}[1]{\oldcal{O}\left(#1\right)}
\usepackage{hyperref}

\usepackage[colorinlistoftodos]{todonotes}

\begin{document}

\title{Six-loop $\varepsilon$ expansion study of three-dimensional $n$-vector model with cubic anisotropy}

\author[label1]{L.\,Ts.\,Adzhemyan}
\author[label1]{E.\,V.\,Ivanova}
\author[label1]{M.\,V.\,Kompaniets}
\author[label1]{\corref{cor2}A.\,Kudlis}
\ead{andrewkudlis@gmail.com}\cortext[cor2]{Corresponding author}

\author[label1]{A.\,I.\,Sokolov}

\address[label1]{St. Petersburg State University, 7/9 Universitetskaya nab., St. Petersburg, 199034 Russia}
%\address[label2]{ITMO University, Kronverkskii ave 49, St. Petersburg, 197101, Russia}

\date{\today}

\begin{abstract}
The six-loop expansions of the renormalization-group functions of 
$\varphi^4$ $n$-vector model with cubic anisotropy are calculated within the minimal subtraction (MS) scheme in $4 - \varepsilon$ dimensions. The $\varepsilon$ expansions for the cubic fixed point coordinates, critical exponents corresponding to the cubic universality class and  marginal order parameter dimensionality $n_c$ separating different regimes of critical behavior are presented. Since the $\varepsilon$ expansions are divergent numerical estimates of the quantities of interest are obtained employing proper resummation techniques. The numbers found are compared with their counterparts obtained earlier within various field-theoretical approaches and by lattice calculations. In particular, our analysis of $n_c$ strengthens the existing arguments in favor of stability of the cubic fixed point in the physical case $n = 3$. 
\end{abstract}

\begin{keyword}
renormalization group, cubic anisotropy, multi-loop calculations, $\varepsilon$ expansion, critical exponents.

\MSC{82B28}
\end{keyword}
\maketitle

\section{Introduction}
As is well known, the systems undergoing continuous phase transitions demonstrate the universal critical behavior. This leads to the concept of classes of universality introduced decades ago. They are determined by the general properties of the system such as spatial dimensionality, symmetry, and the number of order parameter components, thereby its microscopic nature does not play any role in the vicinity of phase transition temperature. There is a set of universal parameters such as critical exponents, critical amplitude ratios, etc. that characterize the critical behavior of the systems belonging to the same universality class. 

The analysis of critical phenomena in a broad variety of materials can be performed on the base of three-dimensional $O(n)$-symmetric $\varphi^4$ field model. In case of one-component -- scalar -- order parameter ($n = 1$) one deals with the Ising model describing phase transitions in uniaxial ferromagnets, simple fluids, binary mixtures, and many other systems. There is also a great numbers of substances with the vector ordering, e.g. easy-plane ferromagnets, superconductors and superfluid helium-4 ($n = 2$), Heisenberg ferromagnets ($n = 3$), quark-gluon plasma in some models of quantum chromodynamics $(n = 4)$, superfluid helium-3 ($n = 18$) and the neutron star matter ($n = 10$). On the other hand, if we consider real materials with more or less complex structure, some anisotropy of the order parameter often exists. Perhaps, simplest example of such a material is a cubic ferromagnet.

Initially, to describe its thermodynamics near Curie point the $O(3)$-symmetric theory neglecting crystal anisotropy has been used. The detailed analysis performed later within the renormalization-group (RG) approach has shown, however, that for proper description of the critical behavior of real cubic crystals one should take into account the presence of the anisotropy, i. e. add to the Landau-Wilson Hamiltonian an extra term invariant with respect to the cubic group of transformations. It looks as $g_2 \sum_{\alpha=1}^{n}\varphi^4_\alpha$, where $\varphi_\alpha$ is $n$-vector ordering field and $g_2$ -- anisotropic coupling constant. This new quartic coupling, in particular, accounts for the fact that in real ferromagnets ($n = 3$) the vector of magnetization "feels" the crystal anisotropy and can lie only along the axes or spatial diagonals of cubic unit cell in the ordered phase. 

\begin{figure}[h!]
\centering
\includegraphics[width=6cm]{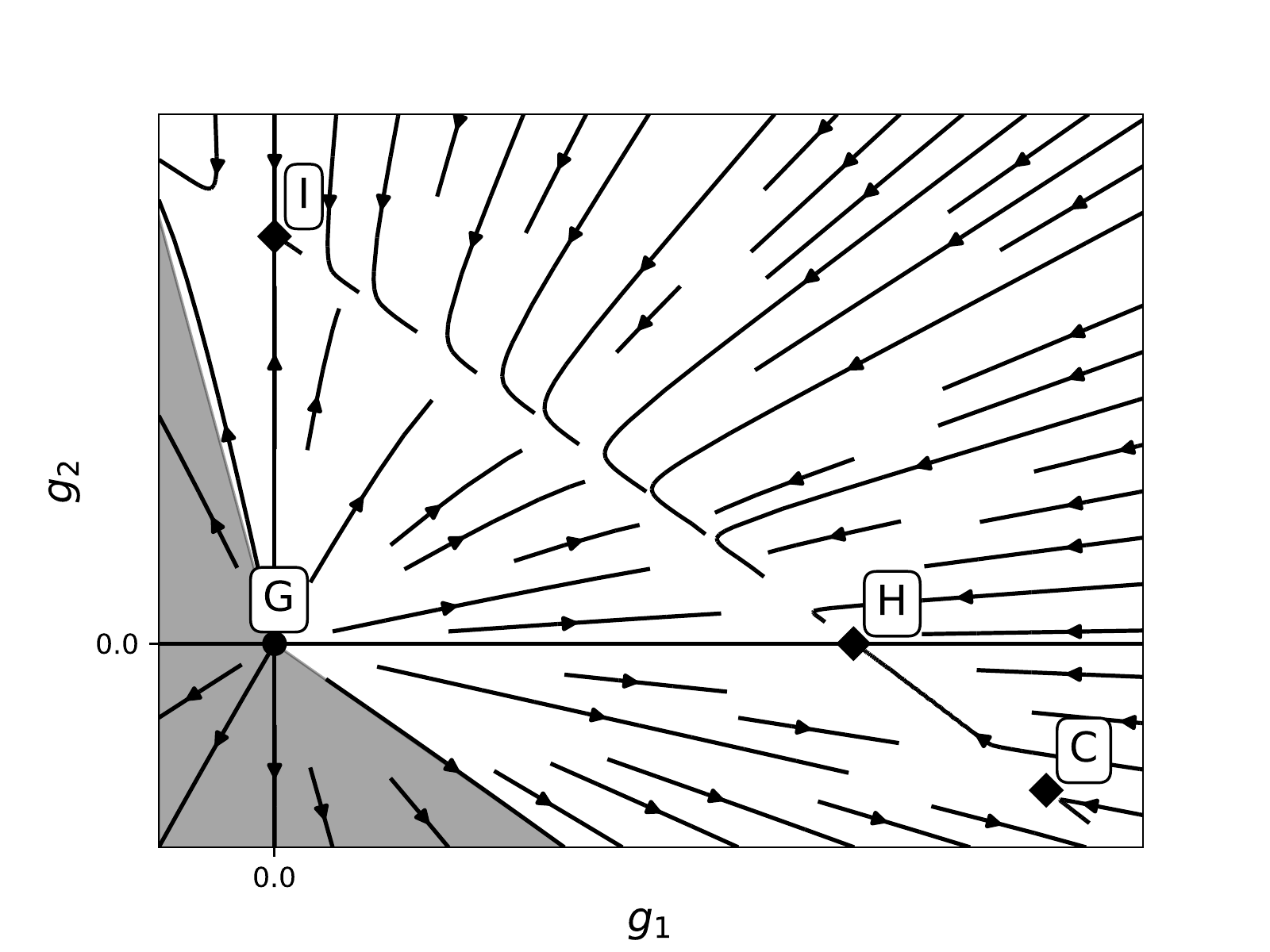}
\includegraphics[width=6cm]{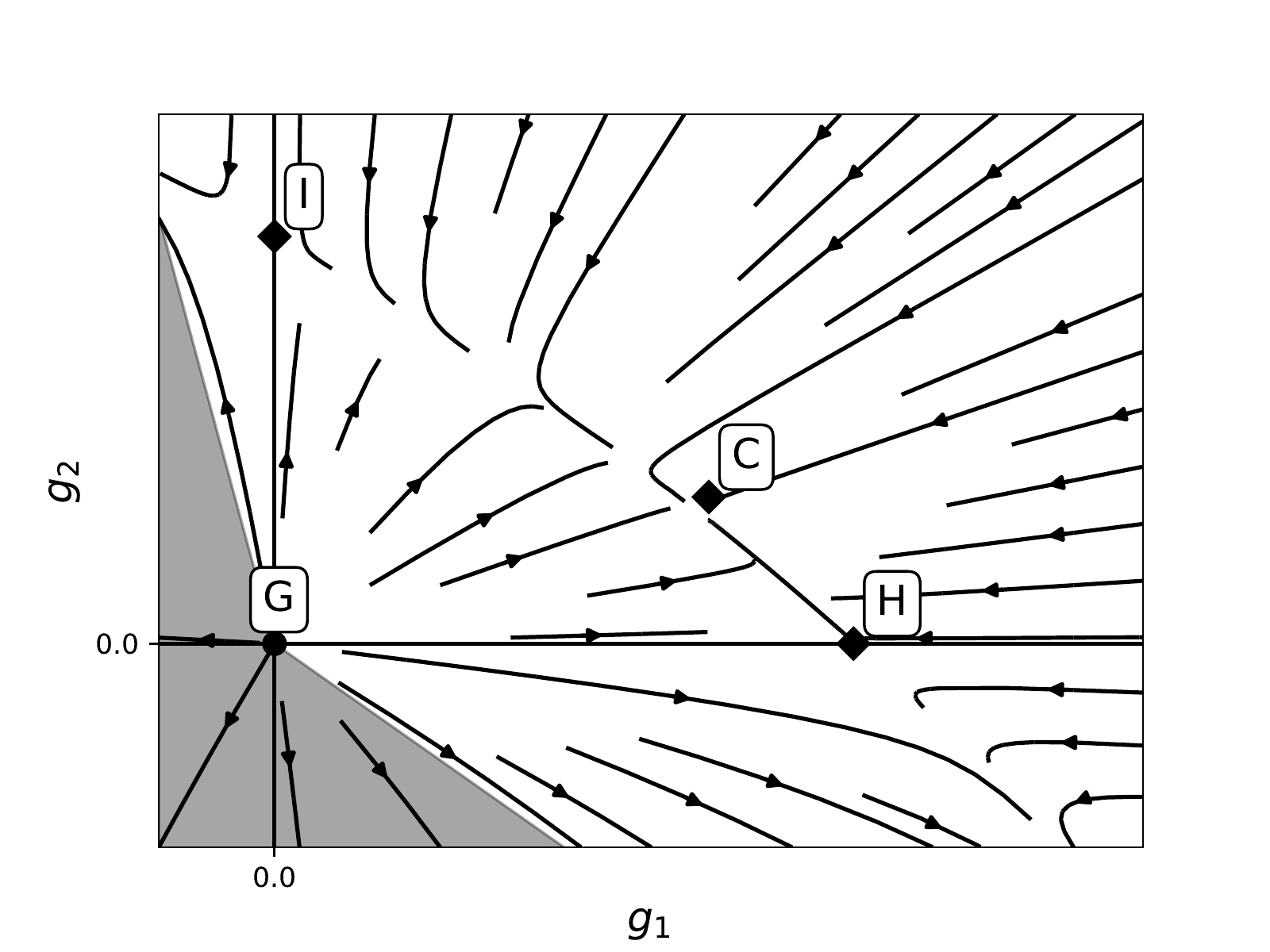}
\caption{RG flows of renormalized coupling constants. The left picture corresponds to $n < n_c$, the right one -- to $n > n_c$. Symbols in boxes mark Gaussian, Ising, Heisenberg and cubic fixed points.} 
\label{fig:rgflow}
\end{figure}

This model with two coupling constants -- $g_1$ (isotropic) and $g_2$ -- was carefully examined since 1972 \cite{WF72} by many researches. As was found, its RG equations describing evolution of quartic couplings under $T\to T_c$ possess four fixed points: Gaussian $(0,0)$, Ising $(0,g_I^*)$, Heisenberg $(g_H^*,0)$ and cubic$(g_{1}^*,g_2^*)$. One of the most important issues involved in the study is the determination of the stability of these fixed points or, in other words, what critical regime takes place in real ferromagnets. Analyzing the RG flows it was shown that the first two points are always unstable for arbitrary values of order parameter dimensionality $n$ whereas the last two of them corresponding to the Heisenberg (isotropic) and cubic (anisotropic) modes of critical behavior compete with each other. Which regime turns out to be stable depends on $n$. For $n < n_c$, where $n_c$ is some marginal value of spin dimensionality, the isotropic (Heisenberg) critical regime is stable while for $n > n_c$ the cubic critical behavior is realized. If initial ("bare") values of coupling constants lie outside the regions of fixed points attraction critical fluctuations strongly modify the behavior of the system converting the second-order phase transition into the first-order one. Figure \ref{fig:rgflow} illustrates the situation. 
 
Thus, in the case $n > n_c$ the cubic quartic term is certainly relevant and has to be taken into account. This results in the emergence of new class of universality corresponding to the anisotropic -- cubic -- critical behavior. So, the value of $n_c$ becomes of prime physical importance since it determines the true regime of the critical behavior in real cubic ferromagnets and of some other systems of interest.

Detailed study of the $n$-vector cubic model including evaluation of critical exponents and $n_c$ was carried out by many groups \cite{AAh73,W73,KW73,NK74,BGZ74,LP75,NT75,TN76,K76,YH77,S77,FDC81,NR82,MS87,MS88,MSS89,N89,KT95,KS95,SAS97,KTS97,CH98,PS99,V00,CPV00,FH00,FHY00,PS01,TMVD02,HV11,KS16} having used both field-theoretical methods and lattice calculations. Early numerical estimates of $n_c$ obtained in the lower-order approximations within the $\varepsilon$ expansion approach \cite{AAh73,KW73,NK74,BGZ74} and in the frame of 3D RG machinery \cite{S77,MS87,MS88} turned out to be in favor of the conclusion that $n_c > 3$, while lattice calculations implied $n_c$ is practically equal to 3 \cite{FDC81}. This made the study of the cubic class of universality less interesting from the physical point of view. Later, however, the higher-order analysis including resummation of RG perturbative series was performed and shown that numerical value of $n_c$ falls below 3 \cite{MSS89,N89,KT95,KS95,SAS97,KTS97,PS99,V00,CPV00,FHY00,KS16}. To date, the most advanced estimates of $n_c$ obtained within the $\varepsilon$ expansion, 3D RG and pseudo-$\varepsilon$ expansion approaches are $n_c = 2.855, 2.87$ \cite{SAS97,CPV00}, $n_c = 2.89, 2.91$ \cite{PS99,CPV00} and $n_c = 2.86$ \cite{FHY00,KS16}, respectively.

These numbers differ from each other appreciably what may be considered as a stimulus to find the value of $n_c$ with higher accuracy. On the other hand, recently the $\varepsilon$ expansions of record length -- six-loop -- for $O(n)$-symmetric $\varphi^4$ field theory \cite{BCK16,KompanietsPanzer:LL2016,KP17} were calculated. This paves the way to analysis of the critical behavior of the cubic model within the highest-order $\varepsilon$ approximation including getting precise numerical estimates for critical exponents and $n_c$. Such an analysis is the aim of this work.    

The paper is organized as follows. In Sec. 2 we write down the fluctuation Hamiltonian (Landau-Wilson action) of $n$-vector cubic model and describe the renormalization procedure. In Sec. 3 the six-loop $\varepsilon$ expansions for $\beta$ functions, critical exponents and $n_c$ are calculated. The six-loop $\varepsilon$ series for cubic fixed point coordinates and critical exponents are also presented here for the physically interesting case $n = 3$. In Sec. 4 the $\varepsilon$ expansions for "observables" -- $n_c$ and critical exponents -- are resummed and corresponding numerical estimates are found. In Sec. 5 the numbers obtained are discussed and compared with their counterparts given by alternative field-theoretical approaches and extracted from the lower-order approximations. Sec. 6 contains the summary of main results and concluding remarks. 
 
\section{Model and renormalization}

In this work we address the field-theoretical RG approach in spatial dimensionality $D = 4-\varepsilon$.\footnote{Original six-loop calculations \cite{BCK16,KompanietsPanzer:LL2016,KP17} were performed in space dimension $D=4-2\varepsilon$ which is more common for high energy physics.} The critical behavior of the cubic model is governed by the well-known Landau-Wilson action with two coupling constants
\begin{equation}
S = \int d^D{x}\Biggl\{\frac{1}{2} \left[(\partial \varphi_{0\alpha }) ^2+ m_0^2  \varphi_{0\alpha }^2\right] + \frac{1}{4!}\left[ g_{01}T^{(1)}_{\alpha\beta\gamma\delta} +g_{02}T^{(2)}_{\alpha\beta\gamma\delta} \right] \varphi_{0\alpha} \varphi_{0\beta}  \varphi_{0\gamma} \varphi_{0\delta} \Biggr\}, \label{H}
\end{equation}
where $\varphi_{0\alpha}$ is $n$-component bare field, $g_{01}$ and $g_{02}$ being the bare coupling constants. The tensor factors $T^{(1)}$ and $T^{(2)}$ entering the $O(n)$-invariant and cubic terms respectively are as follows
\begin{equation}
\begin{split}
T^{(1)}_{\alpha\beta\gamma\delta} = \frac{1}{3}(\delta_{\alpha \beta} \delta_{\gamma \delta} +\delta_{\alpha \gamma} \delta_{\beta \delta}+\delta_{\alpha \delta} \delta_{\gamma \beta}),\\
T^{(2)}_{\alpha\beta\gamma\delta}= \delta_{\alpha\beta\gamma\delta},\quad \delta_{\alpha_1 \dots \alpha_n} = 
\begin{cases}
   1, & \alpha_1= \alpha_2 =\ldots =\alpha_n \\
   0, &\text{otherwise}.
 \end{cases}
 \end{split}
 \label{T12}
 \end{equation}
 In particular, 
 \begin{equation}
 T^{(1)}_{\alpha\beta\gamma\delta}T^{(1)}_{\alpha\beta\gamma\delta} = \frac{n(n+2)}{3}, \qquad T^{(1)}_{\alpha\beta\gamma\delta}T^{(2)}_{\alpha\beta\gamma\delta} = n, \qquad T^{(2)}_{\alpha\beta\gamma\delta}T^{(2)}_{\alpha\beta\gamma\delta} = n.  
\label{3}
\end{equation}
The action (1) is seen to be physical (positively defined) if $g_{02} > - g_{01}$ for $g_{01} > 0$ and $g_{02} > -n g_{01}$ for negative $g_{01}$.

The model is known to be multiplicatively renormalizable. The bare parameters $g_{10},g_{20}, m_0^2, \varphi_0$ can be expressed via the renormalized ones  $g_{1},g_{2}, m^2, \varphi$ by means of the following relations
\begin{equation}
\begin{split}
m_0^2= m^2 Z_{m^2}, \qquad  g_{01} = g_1 \mu^{\varepsilon}Z_{g_1}, \qquad g_{02} = g_2 \mu^{\varepsilon}Z_{g_2}, \qquad \varphi_0 = \varphi Z_{\varphi} ,\\
Z_1=Z_\varphi^2 ,  \qquad      Z_2=Z_{m^2} Z_\varphi^2, \qquad Z_3=Z_{g_1} Z_\varphi^4 ,\qquad Z_4=Z_{g_2} Z_\varphi^4. 
\end{split}
\end{equation}
Using these relations we arrive to the renormalized action 
\begin{equation}
S^R = \int d^D{x}\; \Biggl\{\frac{1}{2} \left[Z_1(\partial \varphi_{\alpha} )^2+ Z_2 m^2 \varphi_{\alpha}^2\right] + \frac{1}{4!}\left[Z_3 g_{1}\mu^{\varepsilon}\; T^{(1)}_{\alpha\beta\gamma\delta} +Z_4 g_{2}\mu^{\varepsilon}\; T^{(2)}_{\alpha\beta\gamma\delta} \right] \varphi_\alpha \varphi_\beta  \varphi_\gamma \varphi_\delta \Biggr\},
\end{equation}
where $\mu$ is an arbitrary mass scale introduced to make couplings $g_1$ and $g_2$ dimensionless. Renormalization constants are defined in a way enabling to absorb divergences from all Green functions, so that renormalized Green functions are free of divergences. Due to multiplicative renormalizability of the model it is enough to remove divergences in two- and four-point one-particle irreducible Green functions:
\begin{equation}
\Gamma^{(2)}_{\alpha\beta} = \Gamma^{(2)}\delta_{\alpha \beta}, \qquad \Gamma^{(4)}_{\alpha\beta\gamma\delta} = \Gamma^{(4)}_1 T^{(1)}_{\alpha\beta\gamma\delta}+\Gamma^{(4)}_2 T^{(2)}_{\alpha\beta\gamma\delta},
\end{equation} 
\begin{equation}
\Gamma_1^{(4)} = \frac{3 (T^{(1)}_{\alpha\beta\gamma\delta}-T^{(2)}_{\alpha\beta\gamma\delta})}{n(n-1) } \Gamma^{(4)}_{\alpha\beta\gamma\delta}, \qquad
\Gamma_2^{(4)} = \frac{(n+2) T^{(2)}_{\alpha\beta\gamma\delta}-3 T^{(1)}_{\alpha\beta\gamma\delta}}{n(n-1)} \Gamma^{(4)}_{\alpha\beta\gamma\delta}.
\label{projectors}
\end{equation} 

In this paper we employ the Minimal Subtraction (MS) scheme where renormalization constants acquire only pole contributions in $\varepsilon$ and depend only on $\varepsilon$ and coupling constants:
\begin{equation}
Z_i(g_1,g_2,\varepsilon) = 1+\sum_{k =1}^{\infty} Z_i^{(k)}(g_1, g_2)\;\varepsilon^{-k}.
\label{Zi}
\end{equation}
Renormalization constants can be found from the requirement of the finiteness of renormalized two- and four-point one-particle irreducible Green functions. Another way to calculate renormalization constants is use of Bogolubov-Parasiuk $R'$ operation:
\begin{equation}
Z_i=1 + KR' \bar{\Gamma}_i,
\end{equation} 
where $R'$ -- incomplete Bogoludov-Parasiuk $R$-operation, $K$ -- projector of the singular part of the diagram and $\bar \Gamma_i$ -- normalized Green functions of the basic theory (see e.g. \cite{Vasilev,BogShirk}) defined by the following relations:
\begin{equation}
\bar{\Gamma}_1=\frac{\partial}{\partial {m^2}}\Gamma^{(2)}\mid_{p=0},  \quad \bar{\Gamma}_2=\frac{1}{2}\left(\frac{\partial}{\partial p}\right)^2\Gamma^{(2)}\mid_{p=0}  \quad \bar{\Gamma}_3=\frac{1}{g_1 \mu^{\varepsilon}}\Gamma^{(4)}_1\mid_{p=0}, \quad 
\bar{\Gamma}_4=\frac{1}{ g_2 \mu^{\varepsilon}}\Gamma^{(4)}_2\mid_{p=0}\;.
\end{equation}

One of the most important advantages of the Bogoludov-Parasiuk approach is that counterterms of the diagrams computed for O(1)-symmetric (scalar) model can be easily generalized to any theory with non-trivial  symmetry due to the factorization of the tensor structures (see e.g. \cite{antonov2013critical,antonov2017critical,KalagovKompanietsNalimov:U(r)}). To calculate tensor factors for particular diagrams of the cubic model \eqref{H} one should apply projectors \eqref{projectors} to it. Such an operation can be automated with FORM \cite{Vermaseren:NewFORM} and GraphState \cite{BatkovichKirienkoKompanietsNovikov:GraphState} while counterterm values can be taken from data obtained in the course of recent 6-loop calculations for $O(n)$-symmetric model~\cite{KP17}.

\section{Six-loop expansions for RG functions, cubic fixed point coordinates, critical exponents and $n_c$}

The RG functions, i. e.  $\beta$ functions and anomalous dimensions $\gamma_{\varphi}$, $\gamma_{m^2}$ are related to renormalization constants $Z_i$ by the following relations: 
\begin{equation}
\begin{split}
\beta_i(g_1,g_2,\varepsilon) =\mu \frac{\partial g_i}{\partial\mu} \mid_{g_{01},g_{02}} = -g_i\left[\varepsilon -  g_1 \frac{\partial Z_{g_i}^{(1)}}{\partial {g_1}}-g_2 \frac{\partial Z_{g_i}^{(1)}}{\partial{g_2}}\right], \quad i = 1,2,\\
\gamma_j(g_1,g_2) = \mu \frac{\partial \log Z_j}{\partial \mu}\mid_{g_{01},g_{02}} = -  g_1 \frac{\partial Z_j^{(1)}}{\partial {g_1}}-g_2 \frac{\partial Z_j^{(1)}}{\partial{g_2}} , \quad j = \varphi, m^2,
\end{split}
\label{sl11}
\end{equation}
where $Z^{(1)}_i$ -- coefficients at first pole in $\varepsilon$ from \eqref{Zi}.

We calculated the RG functions as series in renormalized coupling constants up to six-loop order. They are found analytically and presented in Tables 1, 2, 3 and 4 of Supplementary materials (see~\ref{app:suppl}) in the form
\begin{equation}
\beta_i = g_i\left[-\varepsilon + \sum_{l=1}^6 \sum_{k=0}^{l} C^{k,(l-k)}_{\beta_i} g_1^kg_2^{l-k} \right], \quad i = 1,2,\\
\label{12}
\end{equation}
\begin{equation}
\gamma_j = \sum_{l=1}^6 \sum_{k=0}^{l} C^{k,(l-k)}_{\gamma_j} g_1^kg_2^{l-k}, \quad j = \varphi, m^2.
\label{13}
\end{equation}
The critical regimes of the system are controlled by the fixed points $(g_1^*,g_2^*)$ of RG equations that are zeroes of $\beta$ functions:
\begin{equation}
\beta_1(g_1^*,g_2^*,\varepsilon)=0,\qquad \beta_2(g_1^*,g_2^*,\varepsilon)=0.
\label{beta0}
\end{equation}
As was already mentioned, for the model under consideration there are four fixed points: Gaussian $(0,0)$, Ising $(0,g_I^*)$, Heisenberg $(g_H^*,0)$ and cubic $(g_{1}^*,g_2^*)$. Since six-loop $\varepsilon$ expansions analysis of Ising and Heisenberg models have been performed earlier \cite{BCK16,KompanietsPanzer:LL2016,KP17} we concentrate on the cubic critical behavior. To calculate $\varepsilon$ expansions for critical exponents we have to find those for coordinates of the cubic fixed point. 
Solving \eqref{beta0} by means of iterations in $\varepsilon$ for the cubic fixed point we find: 
\begin{equation}
\begin{split}
g_1^* = \frac{\varepsilon}{n} + \varepsilon^2 \biggl(-\frac{106}{27 n^3}+\frac{125}{27 n^2}-\frac{19}{27 n}\biggr) + \sum_{k=3}^{6} C^{(k)}_{g_1} \varepsilon^k + \bigo{\varepsilon^7},\\
g_2^* = \frac{\varepsilon(n-4)}{3n} + \varepsilon^2 \biggl(\frac{424}{81 n^3}-\frac{178}{27 n^2}+\frac{31}{27 n}+\frac{17}{81}\biggr) + \sum_{k=3}^{6} C^{(k)}_{g_2} \varepsilon^k + \bigo{\varepsilon^7},
\end{split}
\label{15}
\end{equation}
where higher-order coefficients $C^{(k)}_{g_1}$, $C^{(k)}_{g_2}$ are presented in Tables 5 and 6 of Supplementary materials (see~\ref{app:suppl}).

To fully characterize the cubic class of universality, we need to calculate the critical exponents $\alpha$, $\beta$, $\gamma$, $\eta$, $\nu$ and $\delta$. They can be expressed via $\gamma_{m^2}^*\equiv\gamma_{m^2}(g_1^*,g_2^*)$ and $\gamma_{\varphi}^*\equiv\gamma_{\varphi}(g_1^*,g_2^*)$ in the following way: 
\begin{gather}
\alpha = 2-\frac{D}{2+\gamma_{m^2}^*} , \qquad \beta  = \frac{D/2 - 1 + \gamma_{\varphi}^*}{2 + \gamma_{m^2}^*} ,\qquad \gamma  = \frac{2 - 2\gamma_\varphi^*}{2 + \gamma_{m^2}^*} , \qquad \eta = 2\gamma_\varphi^*, \nonumber \\ \qquad \nu = \frac{1}{2+\gamma_{m^2}^*}, \qquad \delta  = \frac{D+2-2\gamma_{\varphi}^*}{D-2+2\gamma_{\varphi}^*}. \label{abgd}
\end{gather}
The critical exponents are related to each other by well-known scaling relations and only two of them may be referred to as independent. 

It is instructive to present $\varepsilon$  expansions of cubic fixed point coordinates for physically important case $n=3$. They are as follows: 
\begin{eqnarray}
g_1^*&=&\frac{1}{3}\varepsilon+\frac{98 }{729}\varepsilon^2+  \varepsilon^3 \left[-\frac{28 \, \zeta (3)}{729}-\frac{61975}{708588}\right]\nonumber \\
&& +\varepsilon^4\left[\frac{30308 \, \zeta (3)}{177147}+\frac{2 \zeta (4)}{729}+\frac{200 \, \zeta (5)}{2187}-\frac{48973747}{344373768} \right]+\nonumber \\
&&+ \varepsilon^5 \left[+\frac{54608659 \, \zeta (3)}{114791256}+\frac{101851 \, \zeta (4)}{708588}-\frac{325 \,\zeta (6)}{39366}-\frac{1519 \, \zeta (7)}{6561} - \right. \nonumber \\
&& \left. -\frac{5375 \, \zeta (3)^2}{59049} -\frac{230560093043}{1338925209984}\right]+\varepsilon^6 \left[\frac{24368284757 \, \zeta (3)}{27894275208}+\frac{597666691 \, \zeta (4)}{1721868840}-\right.\nonumber \\
&&-\frac{1112573461 \, \zeta (5)}{645700815}  -\frac{7725253 \, \zeta (6)}{9565938}+ \frac{16586384 \, \zeta (7)}{7971615} +\frac{176698 \, \zeta (8)}{13286025}+ \nonumber\\
&& +\frac{2911136 \, \zeta (9)}{4782969}-\frac{101024906 \, \zeta (3)^2}{215233605}+\frac{14080 \, \zeta (3)^3}{531441}-\frac{28412 \, \zeta (4) \, \zeta (3)}{177147}+ \nonumber \\
&& \left.+\frac{115696 \, \zeta (5) \, \zeta (3)}{177147}+ \frac{90592\, \zeta(3,5)}{4428675}-\frac{20057900878765}{108452942008704}\right]+\bigo{\varepsilon^7},
\end{eqnarray}
\begin{eqnarray}
g_2^*&=&-\frac{1}{9}\varepsilon+\frac{118 }{2187}\varepsilon^2+\varepsilon^3 \left[\frac{435439}{2125764}-\frac{260 \, \zeta (3)}{2187}\right]+ \nonumber \\
&& +\varepsilon^4\left[-\frac{231404 \, \zeta (3)}{531441}-\frac{226 \, \zeta (4)}{2187}+\frac{920 \, \zeta (5)}{2187}+\frac{257911843}{1033121304}\right]+ \nonumber \\
&& +\varepsilon^5 \left[-\frac{291502339 \zeta (3)}{344373768}-\frac{692465 \, \zeta (4)}{2125764}+\frac{760450 \zeta (5)}{531441}+\frac{22925 \, \zeta (6)}{39366}-\frac{31115 \zeta (7)}{19683} + \right.\nonumber \\
&& \left.+\frac{52853 \zeta (3)^2}{177147}+\frac{1077861709331}{4016775629952}\right] +\varepsilon^6 \left[-\frac{547951382833 \, \zeta (3)}{418414128120}-\frac{631200319 \, \zeta (4)}{1033121304}+ \right.\nonumber \\
&& \left.+\frac{1732037966 \, \zeta (5)}{645700815}+\frac{17543357 \, \zeta (6)}{9565938}-\frac{120541604 \, \zeta (7)}{23914845}-\frac{209656711 \, \zeta(8)}{39858075}+\right.\nonumber \\
&&\left.+\frac{86923264 \, \zeta (9)}{14348907}+\frac{880268036 \, \zeta (3)^2}{645700815} +\frac{490496 \, \zeta (3)^3}{1594323}+\frac{1185542 \, \zeta (4) \, \zeta (3)}{2657205}- \right.\nonumber \\
&& \left.-\frac{708704 \, \zeta (5) \, \zeta (3)}{1594323}+\frac{12497456 \, \zeta (3,5)}{13286025}+\frac{436673550255737}{1626794130130560}\right]+\bigo{\varepsilon^7},
\end{eqnarray}
where $\zeta(3,5)$ is double zeta value \cite{KP17}:
\begin{equation}
\zeta(3,5) = \sum\limits_{0<n<m}\frac{1}{n^3 m^5} \simeq 0.037707672985.
\label{MZV35}
\end{equation}
To give an idea about the numerical structure of these expansions we present them also with the coefficients in decimals:
\begin{eqnarray}
&&g_1^*=0.33333 \varepsilon + 0.13443 \varepsilon^2 - 
 0.13363 \varepsilon^3 + 0.16124 \varepsilon^4 - 
 0.43104 \varepsilon^5 + 1.3278 \varepsilon^6+\bigo{\varepsilon^7}, \nonumber \\
&&g_2^*= - 0.11111 \varepsilon + 0.053955 \varepsilon^2 + 
 0.061933 \varepsilon^3 + 0.050592 \varepsilon^4 - 
 0.18841 \varepsilon^5 + 0.95219 \varepsilon^6+\bigo{\varepsilon^7}.  \nonumber \\
 \label{eq:coordinatesn3}
\end{eqnarray}

The character of a fixed point and, in particular, its stability is determined by the eigenvalues $\omega_1$, $\omega_2$ of the matrix
\begin{eqnarray}
\Omega=\begin{pmatrix}
\dfrac{\partial\beta_1(g_1, g_2)}{\partial{g_1} } & \dfrac{\partial\beta_1(g_1, g_2)}{\partial{g_2}}\\[1.4em] \dfrac{\partial\beta_2(g_1, g_2)}{\partial{g_1}} & \dfrac{\partial\beta_2(g_1, g_2)}{\partial{g_2}}
\end{pmatrix}
\end{eqnarray}
taken at $g_1=g_1^*$, $g_2=g_2^*$.  If both eigenvalues are positive the fixed point is stable and describes true critical behavior. At the same time, the roles of $\omega_1$ and $\omega_2$ in governing the cubic critical behavior are quite different. The eigenvalue $\omega_1$ determines the rate of flow to the cubic fixed point along the radial direction in the plane $(g_1, g_2)$, while $\omega_2$ controls approaching this point normally to the radial ray. In particular, when $n \to n_c$ the cubic fixed point tends to coincide with Heisenberg one and $\omega_2$ goes to zero. So, the dependence of $\omega_2$ on $n$ and its numerical value at $n=3$ are essential in the problem we study. That is why here we write down the $\varepsilon$ expansion for $\omega_2$ only. It reads:      
\begin{equation}
\omega_2 = \varepsilon \frac{n-4}{3n} + \varepsilon^2 \frac{(n-1)(-848+660 n+72 n^2 -19 n^3)}{81 n^3 (n+2)} + \sum_{k=3}^{6} C^{(k)}_{\omega_2} \varepsilon^k+\bigo{\varepsilon^7},
\label{omega2}
\end{equation}  
where coefficients $C^{(k)}_{\omega_2}$, along with those for $\omega_1$, are presented in Tables 7 and 8 of Supplementary materials (see~\ref{app:suppl}).

With $\varepsilon$ expansion for $\omega_2$ in hand we can find $\varepsilon$ series for the marginal dimensionality of the fluctuating field $n_c$. It may be extracted from the equation 
\begin{equation}
\omega_2 (n_c, \varepsilon) = 0.
\end{equation} 
Solving it by iterations in $\varepsilon$ we obtain:
\begin{eqnarray}
n_c &=& 4 - 2 \varepsilon + \varepsilon^2 \left[\frac{5 \zeta (3)}{2}-\frac{5}{12}\right] +\varepsilon^3\left[\frac{15 \, \zeta (4)}{8}+\frac{5 \zeta (3)}{8}-\frac{25 \zeta (5)}{3}-\frac{1}{72}\right]+ \nonumber \\
&&+\varepsilon^4\left[\frac{93 \zeta (3)}{128}+\frac{15\, \zeta (4)}{32}-\frac{3155 \zeta (5)}{1728}-\frac{125 \, \zeta (6)}{12}+\frac{11515 \zeta (7)}{384}-\frac{229 \zeta (3)^2}{144}-\frac{1}{384}\right]+ \nonumber \\
&&+\varepsilon^5 \left[\frac{1709 \zeta (3)}{6912}-\frac{2657\, \zeta (3,5)}{160}+\frac{279 \, \zeta (4)}{512}+\frac{4879 \zeta (5)}{20736}-\frac{21175 \, \zeta (6)}{6912}+\frac{182663 \zeta (7)}{41472}+\right. \nonumber \\
&& \left. +\frac{237079 \, \zeta (8)}{2560}-\frac{2554607 \zeta (9)}{23328}-\frac{21685 \zeta (3)^2}{3456}-\frac{1793 \zeta (3)^3}{324}-\frac{229 \, \zeta (4) \zeta (3)}{96}-\right. \nonumber \\
&& \left.-\frac{3455 \zeta (5) \zeta (3)}{216}+\frac{97}{10368} \right]+\bigo{\varepsilon^6}
\end{eqnarray}
or, in decimals,
\begin{eqnarray}
n_c &=& 4 - 2 \varepsilon + 2.588476 \varepsilon^2 - 5.874312 \varepsilon^3 + 16.82704 \varepsilon^4 - 56.62195 \varepsilon^5+\bigo{\varepsilon^6}. 
\label{ncn}
\end{eqnarray}

Six-loop $\varepsilon$ expansions for critical exponents $\eta$ and $\nu$  corresponding to the cubic class of universality result directly from those for anomalous dimensions and scaling relations (\ref{abgd}). In its turn, six-loop $\varepsilon$ expansions for $\gamma_{\varphi}$ and $\gamma_{m^2}$ originate from RG series (\ref{13}) and $\varepsilon$ expansions for the cubic fixed point coordinates. Since $\varepsilon$ expansions for the critical exponents under arbitrary $n$ are extremely lengthy they are presented in Tables 9 and 10 of Supplementary materials (see~\ref{app:suppl}). Here we write down them only for physically interesting case $n = 3$: 
\begin{eqnarray}
\eta &=&\frac{5}{243} \varepsilon^2 +\frac{4433 }{236196}\varepsilon^3+\varepsilon^4 \left[\frac{2102395}{229582512}-\frac{856}{59049}\zeta (3)\right]+\varepsilon^5\left[-\frac{211933 \zeta (3)}{19131876}-\frac{214 \, \zeta (4)}{19683}+\right. \nonumber \\
&& \left. +\frac{880 \zeta (5)}{19683}+\frac{302817233}{223154201664}\right]+\varepsilon^6 \left[-\frac{123938827 \zeta (3)}{55788550416}-\frac{211933 \, \zeta (4)}{25509168} +\right. \nonumber \\
&& \left.+\frac{80933 \zeta (5)}{3188646}+\frac{1100 \, \zeta (6)}{19683}-\frac{80458 \zeta (7)}{531441}+\frac{169100 \zeta (3)^2}{14348907}-\frac{120071712419}{72301961339136} \right] +\bigo{\varepsilon^7}= \nonumber\\ &=& 0.020576 \varepsilon^2+0.018768 \varepsilon^3-0.0082681 \varepsilon^4+0.022634 \varepsilon^5-0.065781 \varepsilon^6 + \bigo{\varepsilon^7},
\label{eta3}
\end{eqnarray}
\begin{eqnarray}
\nu^{-1} &=&2-\frac{4 }{9}\varepsilon-\frac{383}{2187}\varepsilon^2+ \varepsilon^3\left[\frac{400}{2187} \zeta (3)-\frac{181229}{2125764}\right]+ \nonumber \\
&& +\varepsilon^4 \left[\frac{52279 \zeta (3)}{531441}+\frac{100 \, \zeta (4)}{729}-\frac{3760 \zeta (5)}{6561}-\frac{45792931}{2066242608}\right]+ \nonumber  \\
&& +\varepsilon^5 \left[\frac{6730303 \zeta (3)}{172186884}+\frac{52279 \, \zeta (4)}{708588}-\frac{357650 \zeta (5)}{1594323}-\frac{4700 \, \zeta (6)}{6561}+\frac{38710 \zeta (7)}{19683}-\right. \nonumber \\
&&\left. -\frac{20032 \zeta (3)^2}{177147}+\frac{18998350495}{2008387814976}\right]+\varepsilon^6\left[-\frac{12508116067 \zeta (3)}{167365651248}+\frac{6730303 \, \zeta (4)}{229582512}+\right. \nonumber \\
&&\left. +\frac{137705935 \zeta (5)}{1549681956}-\frac{1076375 \, \zeta (6)}{3188646}+\frac{10154279 \zeta (7)}{19131876}+\frac{94237301 \, \zeta (8)}{15943230}- \right. \nonumber \\
&&\left. -\frac{101478944 \zeta (9)}{14348907}-\frac{44681927 \zeta (3)^2}{129140163}-\frac{560896 \zeta (3)^3}{1594323}-\frac{10016 \, \zeta (4) \zeta (3)}{59049}-\right.\nonumber\\&&\left.-\frac{1565872 \zeta (5) \zeta (3)}{1594323}-\frac{2714888 \, \zeta (3,5)}{2657205}+\frac{21979362510179}{650717652052224}\right]+\bigo{\varepsilon^7} = \nonumber\\ &=& 2-0.44444 \varepsilon-0.17513 \varepsilon^2+0.13460 \varepsilon^3-0.34969 \varepsilon^4+\nonumber\\&&\qquad\qquad\qquad\qquad\qquad\qquad\qquad+0.99461 \varepsilon^5-3.48637 \varepsilon^6+\bigo{\varepsilon^7}.
\label{nu3}
\end{eqnarray}

Of significant interest is also the critical exponent of susceptibility $\gamma$ which is usually measured in experiments and extracted from lattice calculations. Coefficients of its $\varepsilon$ expansion at the cubic fixed point under arbitrary $n$ are presented in Table 11 of Supplementary materials (see~\ref{app:suppl}). For $n = 3$ this expansion is as follows:
\begin{eqnarray}
\gamma &=&1 +\varepsilon\frac{2}{9}+\varepsilon^2\frac{277}{2187}+\varepsilon^3\left[-\frac{200\zeta(3)}{2187}+\frac{85931}{1062882}\right]+ \nonumber  \\ \nonumber
&&+\varepsilon^4\left[-\frac{87775\zeta(3)}{1062882}-\frac{50\zeta(4)}{729}+\frac{1880\zeta(5)}{6561}+\frac{23261567}{516560652}\right]+ \\ \nonumber
&&+\varepsilon^5\left[-\frac{10826597\zeta(3)}{172186884}-
\frac{87775\zeta(4)}{1417176}+\frac{346225\zeta (5)}{1594323}+\frac{2350\zeta(6)}{6561}-\frac{19355\zeta(7)}{19683}+\right.\\ &&\left.+\frac{10016 \zeta(3)^2}{177147}+\frac{2452679419}{125524238436}\right]+\varepsilon^6\left[-\frac{384088139 \zeta(3)}{83682825624}-\frac{10826597\zeta(4)}{229582512}-\right. \nonumber  \\
&& +\frac{240030707 \zeta(5)}{3099363912}+\frac{1913375\zeta(6)}{6377292}-\frac{23980511\zeta(7)}{38263752}-\frac{94237301 \zeta(8)}{31886460}\nonumber  \\ \nonumber &&+\frac{50739472 \zeta(9)}{14348907}+\frac{51810395\zeta(3)^2}{258280326}+\frac{5008 \zeta(3)\zeta(4)}{59049}+\frac{782936\zeta(3)\zeta(5)}{1594323} \\ \nonumber
&&\left.+\frac{280448 \zeta(3)^3}{1594323}+\frac{1357444\zeta(3,5)}{2657205}-\frac{323266891181}{81339706506528}\right]+\bigo{\varepsilon^7} = \nonumber\\ 
&=& 1 + 0.22222\varepsilon + 0.12666\varepsilon^2 - \nonumber 
 0.029080\varepsilon^3 + 0.16865\varepsilon^4 + \\  &&\qquad\qquad\qquad\qquad\qquad\qquad\qquad-0.44336\varepsilon^5 + 1.6059\varepsilon^6+\bigo{\varepsilon^7}.
\label{gammaeps}
\end{eqnarray}

All calculated $\varepsilon$ expansions are rather complicated and need to be checked up. We compared them with known five-loop series \cite{KS95} and found complete agreement. In the Ising ($g_1 \to 0$) and Heisenberg ($g_2 \to 0$) limits our $\varepsilon$ expansions are found to reduce to their counterparts for $O(n)$-symmetric model \cite{KP17} under $n = 1$ and arbitrary $n$ respectively. Our $\varepsilon$ expansions should also obey some exact relations appropriate to the cubic model with $n = 2$. Such a system possesses a specific symmetry: if the field
$\varphi_{\alpha}$ undergoes the transformation
\begin{equation}
\varphi_1 \to {\frac{\varphi_1 + \varphi_2}{\sqrt 2}}, \quad \varphi_2 \to
{\frac{\varphi_1 - \varphi_2}{\sqrt 2}},
\label{32}
\end{equation}
the coupling constants are also transformed:
\begin{equation}
g_1 \to g_1 + {\frac{3}{2}} g_2, \quad g_2 \to -g_2,
\label{33}
\end{equation}
but the structure of the action itself remains unchanged \cite{WF72}. Since the RG functions  are completely determined by the structure of the action, the RG equations should be invariant with respect to any transformation conserving this structure \cite{K76}. It means that under the transformation (\ref{33}) the $\beta$ functions should transform in an analogous way while all the observables including critical exponents should be invariant with respect to above replacement (see \cite{K76,PS01,AS94} for details and extra examples). The expansions (\ref{12}) and (\ref{13}) do satisfy these symmetry requirements. Moreover, transformation (\ref{33}) converts the Ising fixed point into cubic one and vice versa making them dual under $n = 2$. Six-loop $\varepsilon$ expansions (\ref{15}) reproduce this duality. 

\section{Resummation and numerical estimates}

With six-loop $\varepsilon$ expansions in hand we can obtain advanced numerical estimates for all the quantities of interest. It is well known that $\varepsilon$ expansions as other field-theoretical perturbative series are divergent and for getting proper numerical results some resummation procedures have to be applied. In this paper we address the methods of resummation based upon Pad\'e approximants [L/M] which are the ratios of polynomials of orders $L$ (numerator) and $M$ (denominator) and Borel-Leroy transformation. The Pad\'e-Borel-Leroy technique enables one to optimize the resummation procedure by tuning the shift parameter $b$ and proved to yield accurate numerical estimates for basic models of phase transitions. Much simpler Pad\'e technique that is certainly less powerful will be also used, mainly in order to clear up to what extent the numerical results depend on the resummation procedure. Note that both approaches do not require a knowledge of higher-order (Lipatov's) asymptotics of the $\varepsilon$ expansions coefficients finding of which is a separate non-trivial problem.         
\subsection{Resummation strategy and error estimation}
\label{sec:resumstrat}
Application of Pad\'e approximants and use of Pad\'e-Borel-Leroy resummation technique are rather straightforward and were described in detail in a good number of papers and books. At the same time, the determination of the final estimate of the quantity to be found and evaluation of corresponding error bar (apparent accuracy) are somewhat ambiguous procedures. The point is that the choice of a subset of approximants which can be accepted as working and used to get the asymptotic or averaged estimate of a given order usually may be tuned within a very wide range what may lead to unreliable (unstable) results and overestimation of the accuracy.

Here we suggest clear and consistent strategy for calculating estimates with Pad\'e approximants and Pad\'e-Borel-Leroy technique which is aimed to yield the stable results and reasonable error estimates from order to order. While finding numerical values of physical quantities with Pad\'e approximants we use the following procedure. To estimate the value in $k$-th order of perturbation theory we take into consideration approximants of $k$ and $k-1$ orders (particular values of $[L/M]$ depend on the observable). The reason of accounting for such a subset is to provide the results stable from order to order while keeping the contribution from $k$-th order dominant. From this set of approximants we exclude "maximally off-diagonal" ones, in particular $[0/M]$ and $[L/0]$ as they are known to possess bad approximating properties. We exclude also approximants which have poles in the interval $\varepsilon \in [0,2\varepsilon_{phys}]$ (in our case $\varepsilon_{phys}=1$). The reason for this is as follows: if there is a pole in $\varepsilon \in [0,\varepsilon_{phys}]$ the approximant simply cannot be used to estimate the value at $\varepsilon_{phys}=1$, but even if the pole lying outside this area is still close to $\varepsilon_{phys}=1$ such an approximant cannot give reliable estimate as unphysical pole contribution dominates in this case. Particular choice of the upper bound ($2\varepsilon_{phys}$), namely multiplier 2 is based on our experience and tries to keep a balance between dropping out unsuitable approximants and keeping a total number of working approximants as large as possible.

To estimate the error bar (apparent accuracy) we consider values given by different approximants as "independent measurements" of the quantity and use $t$-distribution $t_{p,n}$ with $p=0.95$ confidence level, i. e. estimates for the value itself and its error are calculated with the following formulas: 
\begin{equation}
\langle x\rangle =\frac{x_1+\ldots+x_n}{n}, \qquad \Delta x = t_{0.95,n} \sqrt{\frac{ (\langle x\rangle -x_1)^2+\ldots+(\langle x\rangle -x_n)^2}{n (n-1)}}\;.
\label{student}
\end{equation}

In the case of Pad\'e-Borel-Leroy resummation the procedure is almost the same except the fact that we have an additional -- tuning -- parameter $b$. For each particular value of $b$ we perform Borel-Leroy transformation of the original series, construct Pad\'e approximants of $k$ and $k-1$ orders for Borel-Leroy transform and drop out approximants $[0/M]$, $[L/0]$ and those spoiled by pole(s) on positive real axis. To find the optimal value of $b$ we perform discrete scan over $b\in[0,20]$ with $\Delta b = 0.01$ and search for the value of $b$ which minimizes the standard deviation. The final estimate and error bar are then calculated with \eqref{student} for this value of $b$.

\subsection{Marginal field dimensionality $n_c$}

Let us start from the estimation of the fluctuating field marginal dimensionality $n_c$. As seen from (\ref{ncn}) $\varepsilon$ expansion for $n_c$ is alternating and its coefficients rapidly grow in modulo. The former property makes employing Pad\'e approximants not meaningless. The results of Pad\'e resummation of the series (\ref{ncn}) under the physical value $\varepsilon = 1$ are shown in Table \ref{PadenC}.
\begin{table}[h!]
\caption{Pad\'e triangle for the $\varepsilon$ expansion of $n_c$. Here Pad\'e estimate of $k$-th order (lower line, RoC) is the number given by corresponding diagonal approximant [L/L] or by a half of the sum of the values given by approximants [L/L$-$1] and [L$-$1/L] when a diagonal approximant does not exist. Three  estimates are absent because corresponding Pad\'e approximants have poles close to the physical value $\varepsilon = 1$.} 
\label{PadenC}
\renewcommand{\tabcolsep}{0.4cm}
\begin{tabular}{{c}|*{6}{c}}
$M \setminus L$ & 0 & 1 & 2 & 3 & 4 & 5 \\
\hline
0 & 4 & 2 & 4.5885 & $-$1.2858 & 15.5412 & $-$41.0807 \\
1 & 2.6667 & 3.1283 & 2.7917 & 3.0684 & 2.5692 & \text{} \\
2 & - & 2.8930 & 2.9576 & 2.8828 & \text{} & \text{} \\
3 & 1.9518 & - & 2.9138 & \text{} & \text{} & \text{} \\
4 & - & 2.7887 & \text{} & \text{} & \text{} & \text{} \\
5& 0.4549 & \text{} & \text{} & \text{} & \text{} & \text{} \\
\hline 
RoC & 4 & 2.3333 & 3.1283  & 2.8424 & 2.9576 & 2.8983
\end{tabular}
\end{table} 
Applying the procedure described in section~\ref{sec:resumstrat} to the data collected in Table \ref{PadenC} we obtain $n_c^{(4)}=2.9\pm0.4$, $n_c^{(5)}=2.94\pm0.12$ and $n_c^{(6)}=2.89\pm0.14$ as the four-loop, five-loop and six-loop estimates respectively. These estimates are seen to converge to the value close to 2.9 but the rate of convergence and the accuracy are certainly very low. 

\begin{figure}[h]
\centering
\includegraphics[width=12cm]{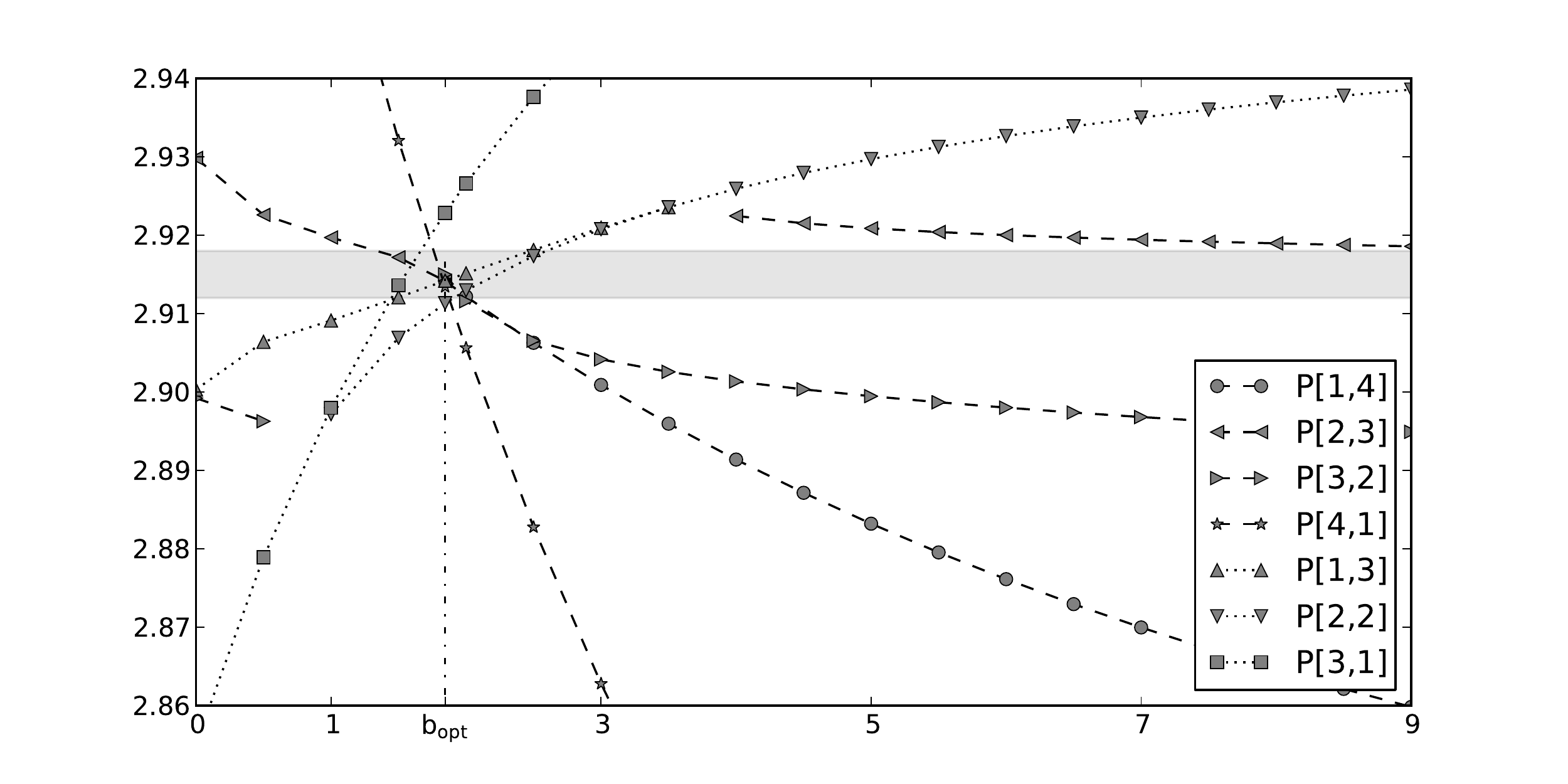}
\caption{Pad\'e-Borel-Leroy estimates of $n_c$ based upon approximants [1/4], [2/3], [3/2], [4/1], [1/3], [2/2] and [3/1] as functions of the parameter b. The curves are depicted only within the intervals where corresponding Pad\'e approximants are free from the "dangerous" (positive axis) poles.}
\label{fig:PBLNc7}
\end{figure}

Since higher-order coefficients of the $\varepsilon$ expansion for $n_c$ are big and rapidly grow use of Borel-Leroy transformation that factorially weakens such a growth should significantly accelerate the convergence and refine the estimate itself. This transformation looks as follows \begin{equation}
f(x) = \sum_{i = 0}^{\infty} c_i x^i = \int\limits_0^{\infty}
e^{-t} t^b F(xt) dt,  \ \
F(y) = \sum_{i = 0}^{\infty} \frac{c_i}{\Gamma(i+b+1)} y^i.
\label{bl}
\end{equation}  
Pad\'e-Borel-Leroy resummation procedure consists of transformation (\ref{bl}) and analytical continuation of the Borel transform $F(y)$ by means of Pad\'e approximants. It includes also the choice (tuning) of the shift parameter $b$ enabling one to achieve the fastest convergence of the iteration scheme. The results of the Pad\'e-Borel-Leroy resummation of the six-loop series for $n_c$ are presented in Fig. \ref{fig:PBLNc7} and Table \ref{PBLnC}. The figure shows the behavior of relevant six-loop and five-loop Pad\'e-Borel-Leroy estimates as functions of the parameter $b$ and illustrates, in particular, the emergence of the  optimal value $b_{opt}$. Note that the curves in Fig. \ref{fig:PBLNc7} are drawn only within the regions where Pad\'e approximants of the Borel-Leroy transform have no positive axis poles. Pad\'e-Borel-Leroy estimates of various approximants obtained under the optimal value of $b$ which was found to be $b_{opt}= 1.845$ are collected in Table \ref{PBLnC}.
\begin{table}[h!]
\caption{Pad\'e-Borel-Leroy estimates of $n_c$ obtained from $\varepsilon$ expansion (\ref{ncn}) under the optimal value of the shift parameter $b_{opt} = 1.845$. The estimate of $k$-th order (lower line, RoC) is the number given by corresponding diagonal approximant [L/L] or by a half of the sum of the values given by approximants [L/L$-$1] and [L$-$1/L] when a diagonal approximant does not exist. Two  estimates are absent because corresponding Pad\'e approximants turn out to be spoiled by dangerous poles.}
\label{PBLnC}
\renewcommand{\tabcolsep}{0.4cm}
\begin{tabular}{{c}|*{6}{c}}
$M \setminus L$ & 0 & 1 & 2 & 3 & 4 & 5 \\
\hline
0 & 4 & 2 & 4.58848 & -1.28584 & 15.5412  & -41.0807 \\
1 & 2.75996 & 3.05988 & 2.87042 & 2.92283 & 2.91341 & \text{} \\
2 & - & 2.93394 & 2.91132 & 2.91499 & \text{} & \text{} \\
3 & 2.57775 & 2.91419 & 2.91416  & \text{} & \text{} & \text{} \\
4 & - & 2.91416 & \text{} & \text{} & \text{} & \text{} \\
5& 2.39138 & \text{} & \text{} & \text{} & \text{} & \text{} \\
\hline 
RoC & 4 & 2.3800 & 3.0599  & 2.9022 & 2.9113 & 2.9146
\end{tabular}
\end{table} 

As is seen the application of Pad\'e-Borel-Leroy machinery indeed makes the iteration faster convergent and corresponding estimates much less oscillating. Being processed according to our strategy (Section~\ref{sec:resumstrat}) the numbers presented in Table \ref{PBLnC} give us $n_c^{(4)}=2.96\pm0.11$, $n_c^{(5)}=2.91\pm0.03$ and $n_c^{(6)}=2.915\pm0.003$ at the four-, five- and six-loop levels. The last, highest-order value
\begin{equation} 
n_c = n_c^{(6)} = 2.915 \pm 0.003
\label{nc} 
\end{equation} 
we accept as a final result of our calculations. 

\subsection{Critical exponents}
 
Since the coordinates of the fixed points depend on the normalization conditions adopted their numerical values being non-universal are not interesting from the physical point of view. That is why further we proceed directly to evaluation of critical exponents characterizing the cubic class of universality at $n=3$. Starting from the six-loop $\varepsilon$ expansions for $\eta$ and $\nu^{-1}$ and using well-known scaling relation we obtain $\varepsilon$ expansions for exponents $\alpha$, $\beta$, $\gamma$, $\nu$ and $\delta$. Then we perform Pad\'e and Pad\'e-Borel-Leroy resummation of all the series in hand. As the Pad\'e-Borel-Leroy resummation procedure turns out to be most effective (regular and fast convergent) for $\beta$ and $\gamma$ we present here details of evaluation of these two exponents. Numerical values of $\beta$ and $\gamma$ obtained within Pad\'e and Pad\'e-Borel-Leroy resummation approaches are collected in Tables \ref{padeb}, \ref{pblb}, \ref{padeg} and \ref{pblg}.          
\begin{table}[h!]
\caption{Pad\'e triangle for the $\varepsilon$ expansion of $\beta$. Five estimates are absent because corresponding Pad\'e approximants have poles lying between $\varepsilon = 0$ and $\varepsilon = 2$.} 
\label{padeb}
\renewcommand{\tabcolsep}{0.4cm}
\begin{tabular}{{c}|*{7}{c}}
$M \setminus L$ & 0 & 1 & 2 & 3 & 4 & 5&6 \\
\hline
0 & 0.5 &  0.3611  & 0.3792 & 0.3421 & 0.4301  & 0.1779& 1.059 \\
1 & 0.3913 & 0.3771 &  0.3670 & 0.3682 & 0.3648 & 0.3740 & \text{}\\
2 & 0.3791 & - & 0.3681 & 0.3673 & 0.3674 & \text{} & \text{} \\
3 & 0.3586 & 0.3715 & - & 0.3674 & \text{} & \text{} & \text{} \\
4 & - & 0.3674 & 0.3693 & \text{} & \text{} & \text{} & \text{} \\
5 & 0.2983 & - & \text{} & \text{} & \text{} & \text{} & \text{} \\
6 & - & \text{} & \text{} & \text{} & \text{} & \text{} & \text{} 
\end{tabular}
\end{table}
\begin{table}[h!]
\caption{Pad\'e-Borel-Leroy estimates of $\beta$ obtained from corresponding $\varepsilon$ expansion under the optimal value of the shift parameter $b_{opt} = 3.460$. Several boxes are empty because of dangerous poles spoiling corresponding Pad\'e-Borel-Leroy approximants.}
\label{pblb}
\renewcommand{\tabcolsep}{0.4cm}
\begin{tabular}{{c}|*{7}{c}}
$M \setminus L$ & 0 & 1 & 2 & 3 & 4 & 5&6 \\
\hline
0 & 0.5 &  0.3611  & 0.3792 & 0.3421 & 0.4301  & 0.1779& 1.059 \\
1 & 0.3952 & 0.3768 &  0.3674 & 0.3674 & 0.3664 & 0.3703 & \text{}\\
2 & 0.3808 & - & 0.3674 & 0.3674 & 0.3672 & \text{} & \text{} \\
3 & 0.3653 & 0.3728 & - & - & \text{} & \text{} & \text{} \\
4 & - & 0.3697 & 0.3690 & \text{} & \text{} & \text{} & \text{} \\
5 & 0.3474 &0.3691& \text{} & \text{} & \text{} & \text{} & \text{} \\
6 & - & \text{} & \text{} & \text{} & \text{} & \text{} & \text{} 
\end{tabular}
\end{table} 
\begin{table}[h!]
\caption{Pad\'e triangle for the $\varepsilon$ expansion of $\gamma$. Five estimates are absent because corresponding Pad\'e approximants have poles close to the physical value $\varepsilon = 1$.} 
\label{padeg}
\renewcommand{\tabcolsep}{0.4cm}
\begin{tabular}{{c}|*{7}{c}}
$M \setminus L$ & 0 & 1 & 2 & 3 & 4 & 5&6 \\
\hline
0 & 1 &  1.2222  &1.3489 & 1.3198 & 1.4885  & 1.0451& 2.651 \\
1 & 1.2857 & - &  1.3252 &1.3446 &  1.3663 & 1.3925 & \text{}\\
2 & 1.4275 & 1.3543 & 1.3939 & 1.3733 & - & \text{} & \text{} \\
3 & 1.2905 & 1.3848 & 1.3770 &  1.3879 & \text{} & \text{} & \text{} \\
4 & - &  1.3754 & 1.3832 & \text{} & \text{} & \text{} & \text{} \\
5 & 0.9069 & - & \text{} & \text{} & \text{} & \text{} & \text{} \\
6 & - & \text{} & \text{} & \text{} & \text{} & \text{} & \text{} 
\end{tabular}
\end{table}
\begin{table}[h!]
\caption{Pad\'e-Borel-Leroy estimates of $\gamma$ obtained from six-loop $\varepsilon$ expansion under the optimal value of the shift parameter $b_{opt} = 0.090$. Empty boxes correspond to the approximants spoiled by dangerous poles.}
\label{pblg}
\renewcommand{\tabcolsep}{0.4cm}
\begin{tabular}{{c}|*{7}{c}}
$M \setminus L$ & 0 & 1 & 2 & 3 & 4 & 5&6 \\
\hline
0 & 1 & 1.2222 & 1.3489 & 1.3198 & 1.4885 & 1.0451 & 2.651 \\
1 & - & - & 1.3263 & 1.3438 & 1.3711 & 1.3825 & \text{} \\
2 & - & - & - & - & - & \text{} & \text{} \\
3 & 1.1893 & 1.3631 & 1.3604 & - & \text{} & \text{} & \text{} \\
4 & - & 1.3605 & 1.3629 & \text{} & \text{} & \text{} & \text{} \\
5 & 1.142 & - & \text{} & \text{} & \text{} & \text{} & \text{} \\
6 & - & \text{} & \text{} & \text{} & \text{} & \text{} & \text{} 
\end{tabular}
\end{table} 
Similar tables were calculated for the exponents $\alpha$, $\delta$, $\eta$ and $\nu$. All the final estimates and error bars obtained according to the resummation  procedure suggested in  Section~\ref{sec:resumstrat} are presented in Table \ref{exponents}.

\begin{table}[h!]
\begin{center}
\caption{The values of critical exponents for the cubic class of universality obtained by means of Pad\'e-Borel-Leroy resummation of the six-loop $\varepsilon$ expansions. Corresponding Pad\'e estimates and the differences between Pad\'e-Borel-Leroy estimates and their Pad\'e counterparts are also presented.}
\label{exponents}
\begin{tabular}{{c}|*{6}{c}}
\hline
n = 3&$\alpha$ & $\beta$ & $\gamma$ & $\delta$ & $\eta$&  $\nu$  \\
\hline
PBL resum.& $-$0.09(9) &0.3684(13) & 1.368(12) & 4.733(4) & 0.036(3) &  0.700(8)   \\
\hline
Pade resum.& $-$0.11(6) &0.368(3) & 1.379(8)& 4.772(17)& 0.038(4) & 0.703(5)  \\
\hline
Difference & 0.02(11) & 0.0004(33) & $-0.011(44)$ & $-0.039(17)$ & $-0.002(5)$ & 0.003(9) \\
\hline
\end{tabular}
\end{center} 
\end{table}

What is the accuracy of numerical results just found? Some idea on the point may be obtained looking at the differences between the Pad\'e-Borel-Leroy  and Pad\'e estimates presented in Table~\ref{exponents}. However, much more definite conclusions concerning an actual accuracy of our calculations can be made on the base of the analysis to what extent the numbers obtained obey exact scaling relations between the  critical exponents. One can combine critical exponents in different ways. We choose the next set of independent relations: 
\begin{gather}
1) \ \frac{\gamma}{\nu(2-\eta)} - 1 = 0 , \qquad 2) \ \frac{2\beta}{\nu(1 + \eta)} -1 = 0 ,\qquad 3) \ \frac{5 - \eta}{\delta(1 + \eta)} - 1= 0, \qquad 4)\ \beta + \frac{\alpha +\gamma}{2} - 1 = 0, 
\label{sclrel}
\end{gather}
that are "normalized to unity" to get the estimates of accuracy more uniform. Since the calculated values of critical exponents are approximate they can not meet the scaling relations precisely and emerging discrepancies may be considered as a measure of achieved accuracy. The discrepancies relevant to scaling relations \eqref{sclrel} along with their error bars originating from the estimates of the critical exponents themselves (Table \ref{exponents}, upper line) are presented in Table \ref{scalingnum}.
\begin{table}[h!]
\begin{center}
\caption{Six-loop estimates of critical exponents versus scaling relations}
\label{scalingnum}
\begin{tabular}{{c}|*{4}{c}}
\hline
Scaling relation: & 1 & 2 & 3 & 4  \\
\hline
Deviation from zero & -0.005(14) &0.016(13) &0.0121(36)   & 0.007(45) \\
\hline
\end{tabular}
\end{center} 
\end{table}
As is seen the deviations from exact scaling relations are small demonstrating the consistency of our approach and indicating that actual computational uncertainty of found numerical estimates is of order of 0.01.   

To finalize this section, in Table~\ref{corexponents} we present, for completeness, the values of correction-to-scaling exponents $\omega_1$ and $\omega_2$ obtained by  resummation of corresponding $\varepsilon$ expansions. Despite the fact that zero lies inside the error bar for $\omega_2$ the median value of this exponent, being very small, turns out to be positive. Moreover, keeping in mind the results of independent evaluation of $n_c$ we may state that the value of $\omega_2$ given by six-loop $\varepsilon$ expansion analysis is certainly positive. More accurate estimates for $\omega_2$ can be obtained within the higher-order (seven-loop, etc.) approximations or by means of more sophisticated resummation procedure such as Borel transformation combined with conformal mapping which will be a subject of a separate paper.

\begin{table}[h!]
\begin{center}
\caption{The values of correction-to-scaling exponents $\omega_1$ and $\omega_2$ for the cubic class of universality obtained by means of Pad\'e-Borel-Leroy resummation of the six-loop $\varepsilon$ expansions. Corresponding Pad\'e estimates and the differences between Pad\'e-Borel-Leroy estimates and their Pad\'e counterparts are also presented.}
\label{corexponents}
\begin{tabular}{{c}|*{6}{c}}
\hline
n = 3&$\omega_1$ & $\omega_2$  \\
\hline
PBL resum.& 0.799(4) &0.005(5)  \\
\hline
Pade resum.& 0.78(11) &0.008(38)  \\
\hline
Difference & 0.02(11) & $-0.003(38)$ \\
\hline
\end{tabular}
\end{center} 
\end{table}

\section{Discussion}

In this section we will compare our results with those obtained earlier within the lower-order approximations and by alternative methods.

The first quantity of interest is the marginal spin dimensionality for which we get the value $n_c=2.915(3)$. It is worthy to note that the $\varepsilon$ expansion for this quantity has rapidly growing coefficients (see eq.~\eqref{ncn}) what prevents Pad\'e approximants from giving accurate enough numerical results while Pad\'e-Borel-Leroy approach yields stable estimates with an accuracy increasing from order to order. The results of previous studies performed within the $\varepsilon$ expansion approach and RG machinery in fixed dimensions (3D RG) as well as the numbers extracted from the Monte Carlo simulations and the six-loop pseudo-$\varepsilon$ expansion are aggregated in the Table~\ref{marginaldimall}.

\begin{figure}[h!]
\centering
\includegraphics[width=14cm]{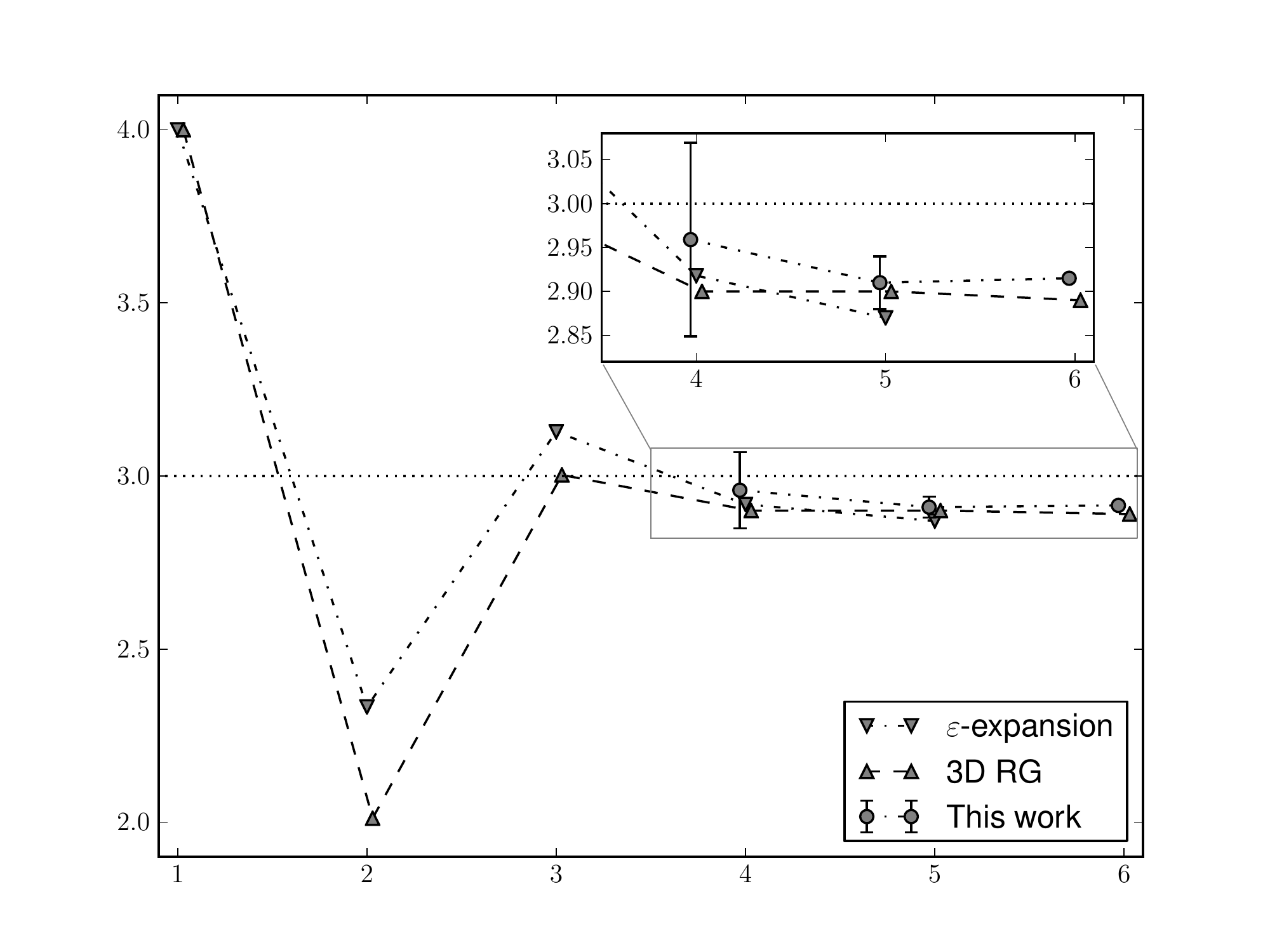}
\caption{Dependence of the marginal spin dimensionality value on the order of RG approximation. The upper curve ("$\varepsilon$~expansion") represents the estimates obtained earlier from the five-loop $\varepsilon$ expansion for $n_c$.} 
\label{fig:trendnc}
\end{figure}

\begin{table}[h!]
\centering
\caption{Marginal order parameter dimensionality $n_c$ given by the $\varepsilon$ expansion technique, 3D RG approach, Monte Carlo simulations and the pseudo-$\varepsilon$ expansion machinery. By the number of loops we mean the order of approximation.}
\label{marginaldimall}
\begin{tabular}{c|cc|cc||cc}
\hline
Number &&&&&& \\ of loops&  $n_c$&Paper   & $n_c$ & Paper & $n_c$ & Paper \\ \hline 
     &\multicolumn{2}{c|}{$\varepsilon$ expansion}   &\multicolumn{2}{c||}{3D RG}&\multicolumn{2}{c}{Others} \\ \hline
1     		 &  4 &\cite{NK74}-1974    		  & 			   &  &\multicolumn{2}{c}{Monte Carlo}\\ \hline
2     		 & 2.333  &\cite{NK74}-1974       & 2.0114    &\cite{J83}-1983   &3  & \cite{CH98}-1998 \\ \hline 
3     		 & 3.128  &\cite{NK74}-1974       & 3.003  &\cite{MS84}-1984    &&\\ \hline
4     		 &  2.918&\cite{SAS97}-1997      & 2.9  &\cite{MSS89}-1989      &&\\ 
             &    2.96(11)     &This work-2019           &   2.89(2) &\cite{V00} -2000         & &\\ \hline
	 		 &	 2.958&  \cite{KS95}-1995  	  &  & &&\\
5     		 &   $<$3& \cite{KTS97}-1997     &  2.89-2.92&\cite{PS99}-2000&&  \\
      		 &   2.855 &  \cite{SAS97}-1997     &  				   &  	 &&\\
      		 &    2.87(5) & \cite{CPV00}-2000     & 				   &   	 && \\
      		 &  	2.91(3) 	& This work-2019   &  			   	   & 	& 
             \multicolumn{2}{c}{Pseudo-$\varepsilon$ expansion} \\  \hline
6     		 &   2.915(3)  & This work-2019   &  2.89(4) &\cite{CPV00}-2000  &  2.86(1) & \cite{KS16}-2016 \\ 
			&   			&				&			&					&	 2.862(5)&		\cite{FHY00}-2000	\\ \hline	
\end{tabular}
\end{table}
\noindent In addition, the values of $n_c$ collected in Table~\ref{marginaldimall} are depicted at Fig. \ref{fig:trendnc} to visualize the trend these values demonstrate under increasing order of approximation. This trend enable us to conclude that $n_c$ is certainly less than 3 for the 3D cubic model that justifies the significance of studying the cubic class of universality.

The other quantities of prime physical importance are critical exponents of the cubic universality class. We should stress that to get estimates for critical exponents we perform resummation of the series for each exponent separately and afterwards checked a validity of several scaling relations \eqref{sclrel}. Despite the fact that sometimes the relations are satisfied with inaccuracies exceeding corresponding error bar estimates, these deviations are not too large lying within 3$\sigma$ interval. This may be considered as a proof of the consistency of the results obtained and a demonstration of the numerical power of the $\varepsilon$ expansion approach. 

It is worthy to compare our estimates with their  analogs given by the lower-order approximations and with the results of multi-loop 3D RG analysis. The data enabling one to do such a comparison are collected in Table \ref{tab:etanu}. The numbers presented in both columns are seen to rapidly converge to the asymptotic values that differ from each other only tiny coinciding in fact within the  declared error bars. It confirms the conclusion that the field theory is a powerful instrument enabling one to get precise numerical results provided the calculations are performed in high enough pertubative order. On the other hand, addressing the six-loop $\varepsilon$ approximation shifts the estimates only slightly indicating that they should be very close to the exact values still unknown.             
\vspace{1cm}
\begin{table}[h]
\caption{Critical exponent values given by multi-loop $\varepsilon$ expansion calculations versus those resulting from 3D RG analysis. Error bar for four loop  estimate of $\eta$ is absent because it  can not be evaluated within approach described in Sec.~\ref{sec:resumstrat}.}
\label{tab:etanu}
\centering
\begin{tabular}{c|ccc|ccc}
\hline
Number &&&&&&\\ of loops             & $\eta$   & $\nu$ &  Paper    & $\eta$    &$\nu$ & Paper \\ \hline 
     &\multicolumn{3}{c|}{$\varepsilon$ expansion}   &\multicolumn{3}{c}{3D RG} \\ \hline
3    &   &       &           &--         &0.700  & \cite{MS84}-1984    \\ \hline
%3    &\cite{NK74}-1974   & --       & --       & \cite{MS84}-1984   &--         &0.700      \\ \hline
4    & 0.034&   0.68(3)     &  This work-2019       & 0.0331    & 0.6944  & \cite{MSS89}-1989   \\
%4    &\cite{SAS97}-1997  & --       & --       & \cite{MSS89}-1989  & 0.0331    & 0.6944    \\
     &                   &          &            & 0.0332    &  0.6996  & \cite{V00} -2000  \\ \hline
5       & 0.0375(5)&0.6997(24)&\cite{MV98}-1998  & 0.025(10) & 0.671(5)& \cite{PS99}-2000   \\ 
       &0.0374(22)& 0.701(4)&\cite{CPV00}-2000 &                    &           &           \\ 
       &0.0353(21)  & 0.686(13)&This work-2019&&&											\\ \hline
6          & 0.036(3) &  0.700(8)& This work-2019         & 0.0333(26)& 0.706(6) & \cite{CPV00}-2000 \\ \hline
\end{tabular}
\end{table}

Another point to be discussed is to what extent -- quantitatively -- the critical exponents of the cubic class of universality differ from those of the 3D Heisenberg model. Since for $n=3$ the cubic fixed point lies near the Heisenberg one corresponding differences are known to be rather small. In Table \ref{comwithheis} we present the estimates of critical exponents for cubic and Heisenberg  
classes of universality obtained in the six-loop approximation. As expected, the differences between numerical values of critical exponents for these two classes are really small. So, it is hardly believed that measuring critical exponents in physical or computer experiments one can distinguish between cubic and Heisenberg critical behaviors.
\begin{table}[h!]
\begin{center}
\caption{Comparison of critical exponents for   cubic (this work) and Heisenberg (\cite{KP17}) classes of universality for $n=3$. The numbers with asterisk were obtained from six-loop $\varepsilon$ expansion estimates for $\eta$ and $\nu$ via scaling relations.}
\label{comwithheis}
\begin{tabular}{{c}|*{6}{c}}
\hline
n = 3&$\alpha$ & $\beta$ & $\gamma$ & $\delta$ & $\eta$&  $\nu$  \\
\hline
Cubic& $-$0.09(9) &0.3684(13) & 1.368(12) & 4.733(4) & 0.036(3) &  0.700(8)   \\
\hline
Heisenberg&$-$0.118(6)*&0.3663(12)*&1.385(4)*&4.781(3)*&0.0378(5)
&0.7059(20)\\
\hline
\end{tabular}
\end{center}
\end{table}

\section*{Conclusion}

To summarize, we performed six-loop RG analysis of the critical behavior of $n$-vector $\varphi^4$ model with cubic anisotropy in the framework of $\varepsilon$ expansion approach employing the minimal subtraction scheme. We calculated $\varepsilon$ expansions for marginal spin dimensionality $n_c$ and  critical exponents $\alpha$, $\beta$, $\gamma$, $\delta$, $\eta$, $\nu$, $\omega_1$, $\omega_2$ for the cubic class of universality. We resummed these diverging series with Pad\'e approximants and using Pad\'e-Borel-Leroy technique. Obtained numerical estimates for critical exponents turn out to be self-consistent in the sense that they are in accord, within the computational uncertainties, with the scaling relations.  Six-loop contributions are found to shift five-loop estimates only slightly but they improve numerical results considerably diminishing their error bars. Our results confirm and strengthen the conclusion that cubic ferromagnets (D = 3, n = 3) belong to cubic class of universality and their critical behavior is described by critical exponents differing from those of 3D Heisenberg model. At the same time, the critical exponents of 3D cubic and Heisenberg models are numerically so close to each other that it makes their behaviors practically indistinguishable if one limits himself by measuring critical exponents only.

\section*{Acknowledgment}
It is a pleasure to thank Professor M. Hnati\v{c} and Professor M.Yu. Nalimov for fruitful discussions. E.I. and A.K. are especially grateful to the Professor Hnati\v{c} for support and hospitality during their stay in Slovakia. This work has been supported by Foundation for the Advancement of Theoretical Physics "BASIS" (grant 18-1-2-43-1).

\appendix
\section{Supplementary materials }
\label{app:suppl}
In Supplementary materials we present expansions of RG functions and critical exponents for arbitrary $n$.
In \textit{rg\_expansion\_coefficients.pdf} we list coefficients $C_k^{i,\,j}$ of the expansions of beta functions $\beta_1$, $\beta_2$ \eqref{12}, anomalous dimensions $\gamma_\phi$, $\gamma_{m^2}$ \eqref{13} and $\varepsilon$ expansions of coordinates of the cubic fixed point \eqref{15}, correction-to-scaling exponents $\omega_1$, $\omega_2$ \eqref{omega2} and critical exponents $\eta$ \eqref{eta3}, $1/\nu$ \eqref{nu3} and $\gamma$ \eqref{gammaeps} corresponding to cubic universality class.

Additionally, for RG functions ($\beta_1(g_1,g_2)$, $\beta_2(g_1,g_2)$, $\gamma_\phi(g_1,g_2)$, $\gamma_{m^2}(g_1,g_2)$) we provide Mathematica file with their expansions (\textit{rg\_expansion.m}).
For critical exponents we present Mathematica files for all non-trivial fixed points: cubic (\textit{cubic\_crit\_exp.m}), Ising (\textit{ising\_crit\_exp.m}) and Heisenberg (\textit{heisenberg\_crit\_exp.m}). Each file contains $\varepsilon$ expansion for exponents $\alpha$, $\beta$, $\gamma$, $\delta$, $\eta$, $\nu$ as well as for $1/\nu$, correction-to-scaling exponents $\omega_1$, $\omega_2$ and coordinates of fixed points $g_1^*$, $g_2^*$. In the file corresponding to cubic fixed point (\textit{cubic\_crit\_exp.m}) we also present expansion for marginal spin dimensionality $n_c$. 
\newpage

\bibliographystyle{elsarticle-num}
\bibliography{cubic}
%%%%%%%%%%%%%%%%%%%%%%%%%%%%
\end{document}